# Development of Highly Efficient Multi-invariable Wireless Sensor System Design for Energy Harvesting

Farid Touati[1,] Usman Asghar[1]
[1]Department of Electrical Engineering, Qatar University,
[1]touatif@qu.edu.qa

Damiano Crescini[2], Alessio Galli[2], Adel Ben Mnaouer[3],
[2]Università di Brescia, Brescia (Italy),
[2]damiano.crescini@ing.unibs.it
[3]Canadian University of Dubai, [2]adel@cud.ac.ae

*Abstract* — *Capillary wireless sensor networks devoted to air quality monitoring have provided vital information on dangerous air conditions. In adopting the environmentally-generated energy as the fundamental energy source the main challenge is the implementation of capillary networks rather than replacing the batteries on a set period of times that leads to functional dilemma of devices management and high costs. In this paper we present a battery-less, self-governing, multi-parametric sensing platform for air quality monitoring that harvests environment energy for long run. Furthermore study on sensor section with their results have also been described in the paper. A customized process of calibration to check the sensors' sensitivity and a basic portfolio of variant energy sources over the power recovery section could productively improve air quality standards tracing in indoor and outdoor application, in a kind of "set and forget" scenario.*

## I. INTRODUCTION (HEADING 1)

Pollutant air has become a very important issue as large cities deal with constant expansion in the number of combustion-engine powered vehicles and in population. In World Health Organization (WHO) figures, seven million people die annually as a result of exposure to air pollution [1]. In every eight deaths globally, one is due to poor air quality, Earth air pollution can be considered as the single biggest environment health risk. Number of studies have shown a correction between poor childhood health and infant fatality in areas with immense concentrations of PM and CO. According to some modern studies the relationship between pollutant air quality and a number of health diseases has been confirmed [2-6]. In the US, around 10-20 million dollars are spent for curing the society in each single year, affected by building disorder diseases [7]. United States Environmental Protection Agency (EPA) has placed poor air quality in category of the five major factors which threatens the human health. Since the indoor air pollution has a large impact on the public health, indoor air quality monitoring has found its place as a cutting-edge smart home application in Internet of Things (IoT) field. According to European Union official facts, about 225000 people die from diseases caused by combustion emissions through cars each year. To fight with this threat, the European Union has introduced rigorous laws and intends to cut down car emissions by 20% by 2020 [8]. These facts conveys that we need to have suitable air pollutants monitoring systems that make the data availability possible in places where traditional monitoring methods are challenging to implement. It also improves the reliability. These requirements led to the design of numerous autonomous systems which enables checking of indoor and outdoor air quality. In the enhancement procedure of air quality (e.g. air sanitation, HVAC system, air cleaning), the predominant objective is to identify correctly the air pollutants and to define the polluted regions so as to provide proper rectification. The advancement in low power micro-electronics and the new inexpensive electrochemical sensors linked to low power wireless techniques have allowed the development of highly efficient, low cost and low power air quality monitoring systems and their deployment in actual environments [9-16]. Advances in wireless connectivity and networking technologies leads to reduced difficulties and installation costs, allowed rapid deployment, remote and easy reconfiguration of air quality monitoring systems. Low cost and autonomous monitoring systems that are able to operate in any kind of environment, especially in severe and turbulent ones, are needed to monitor air quality while reducing the human interference.

There is an elementary problem in deployment of high power-efficiency wireless sensing nodes for long, continuous and autonomous operation. If nodes are operated by battery then the costs of battery replacement will make such systems too overpriced to be deployed in wide area and tough environment. This pattern of independent monitoring systems relies on concept of "set and forget" format, where minimal human intervention is required. This is certainly the system design we are introducing in this paper.

## II. RELATED WORK: AIR POLLUTANT MONITORING SYSTEMS AND POWER SUPPLY PROBLEMS

Number of other research works have already dealt with air quality-monitoring systems based on wireless sensor networked dedicated to diagnosis and data analysis. In [9], the authors considered air quality for outdoor environment in Cincinnati (USA) using as wireless communication system to figure out the concentration level of CO along roads and traffic.

The system comprises of humidity sensor, temperature sensor and a carbon monoxide sensor only. In their study the authors did not consider any other gas sensors or any environmental parameters. Moreover it was based on rechargeable batteries recharged from a single source photovoltaic solar panels. In **[10]**, Devarakonda et al. discussed an air quality monitoring system based on wide wireless GSM network. They deployed their devices in New Jersey (USA) on highways and suburban areas using only a carbon monoxide sensor and attaching a battery on board too. Moreover power sources are not taken into account or studied. The authors of **[11]** conferred an air quality monitoring system targeting both wide area deployment and personal use as well. The system is designed to detect PM2.5. It is interconnected to a wireless network to send data over a distant server. The design was battery-dependent and hence, it may be benefit of power-harvesting add-ons. Jiang et al. in **[12]** have designed, a personal device for indoor air quality monitoring, the system is able to collect data and send them to a Smart-phone through a Wi-Fi network or via Bluetooth. The parameters checked are light intensity, temperature and $CO_2$. This systems sounds attractive for personal use, still concerns are related to the replacement of battery on regular intervals, and the fact that no discussion or mature studies on energy harvesting were presented. Moreover, the number of parameters monitored, some are missing, such as the humidity and some dangerous gases, affecting human health. Another innovative air quality monitoring system based on wireless network have also been reported **[13]**. The recommended design was able to detect $NO_2$, $O_3$, $SO_2$, $NH_3$, and NO and send real-time environmental data over a Wi-Fi Network. The system was also dependent on battery, which needs frequent replacement. In **[14]**, Kim et al. demonstrated an indoor air quality monitoring system built using a Raspberry-pi computer. Their established board included one CO sensor, VOC sensor and $NO_2$ sensor. Again sensors that have been used are known to be power hungry, the board getting power from 5V external power source. None of power harvesting technique were used, thus, a physical deployment on large scale would require regular battery replacement. Honicky et al. presented N/SMARTS **[15]**, which was GPS-enabled, and environmental data acquisition platform that was based on a cell phone. The major transducer module consists of CO and $NO_X$, a three-axis accelerometer, and a temperature sensor using a power source battery pack that demand recharging after some hours of use. Finally, in **[16]** Postolache et al. illustrated a networked air quality monitoring for outdoor use. In this design, every single node is deployed in a tailored scenario that incorporates $SnO_2$ sensor arrays linked to data acquisition card. Looking at the system morphology, in most of cases the nodes are hard-wired to a central control unit. The main drawback of this platform lies with the gas sensors employed, where power dissipation across the electrodes (i.e., 300mW) makes it impossible to be applied in environmentally-powered networks. From the above discussed literature on air quality monitoring structures shows that vital ownership such as estimated power consumption, energy storage, energy harvesting and battery-less permanent operation are not accordingly developed. This paper presents the design, practical implementation and operation of an indoor/outdoor air quality monitoring system based on harvesting environmental-power which enables independent deployment in a "set and forget" method.

## III. THE SYSTEM ARRANGEMENTS

The proposed system (SERENO acronym of SEnsor REceiver NOde) is portrayed in Fig. 1, 2 and 3 in which the most important blocks are shown (Fig. 1 in green blocks related to the power harvesting sub-system, Fig. 2 in red, the blocks showing the six gas sensors and in Fig. 3 in red, the blocks related to some functional parts of the system).

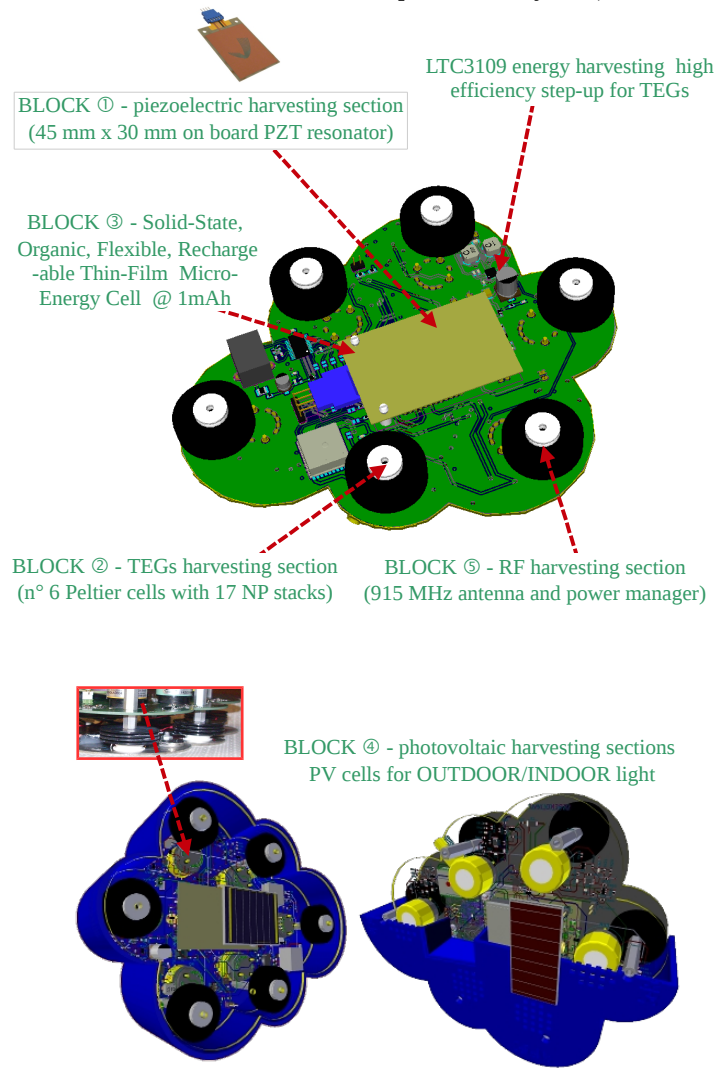

BLOCK ① - piezoelectric harvesting section (45 mm x 30 mm on board PZT resonator)

LTC3109 energy harvesting high efficiency step-up for TEGs

BLOCK ③ - Solid-State, Organic, Flexible, Recharge-able Thin-Film Micro-Energy Cell @ 1mAh

BLOCK ② - TEGs harvesting section (n° 6 Peltier cells with 17 NP stacks)

BLOCK ⑤ - RF harvesting section (915 MHz antenna and power manager)

BLOCK ④ - photovoltaic harvesting sections PV cells for OUTDOOR/INDOOR light

FIGURE 1 – 3D view of SERENO (PCB bottom view) where the harvesting blocks are illustrated in green together with the thin-film battery

## A. Air Quality Sensors for gases concentration

The main characteristics of represented system lies in having an batch of gas sensors able to measure the most effecting pollutants existing in the atmosphere. The research on such kind of products present in the market has led to choose the gas sensors with electrochemical working principal **[17-24]**. Power consumption is not a major issue with them, a reasonably small package size and quick response to the target gas. Other parameters such as good cross-sensitivity to non-target gas elements and slight drift in temperature have been considered. The main factors that differentiate among them bases on the target gases, cost and the resolution of each gas sensor concentration in terms of parts per billion (ppb), coordinating with the EPA rules **[25]**. For our design, we adapted: NE4-CO Carbon monoxide, NE4-$Cl_2$ Chlorine, NE4-$NO_2$ Nitrogen dioxide, NE4-$H_2S$ Hydrogen Sulphide, NE4-$NH_3$ Ammonia sensor and NE4-NO Nitrogen monoxide from Nemoto **[26]** (see Fig.2). The SENSIRION SHT21 humidity and temperature sensor, (Fig. 3) were also adopted because of the fact that the detected information of the gas transducers are sensitized to surrounding temperature and humidity.

## B. Management of Power Consumption and Energy Harvesting Techniques

The developed single board reside of six gas sensors, the temperature and humidity sensors and wireless transmission module with a micro-controller(μC) on single chip, on one board. With regards to the type of applications as a point of reference, whether using a conventional lithium battery or using a solid state battery, it is essential to specify the energy consumption and management techniques employed so as to assure either extended battery power longevity or autonomous energy source ("*set and forget*" scenario).

In the this paper, we are concentrating on the following energy harvesting sources, all these can be active at the same time depending on deployment environment and conditions:

i. *Energy harvesting from vibration (Fig. 1 block ①) devoted to the transformation of otherwise wasted energy from mechanical oscillations into usable electrical energy. The working principal of mechanical resonator is based on a piezoelectric (PZT) material. Examples of generation of power for resonance frequencies of variant combinations are given underneath:*

- Tuned Frequency to 36.5 Hz with 2.27 g seismic mass and acceleration = 0.3 g: Max. power extracted = 0.319 mW(according to Jacobi's law Fig. 4(a))

- *Frequency tuned to 36.5 Hz with 2.27 g seismic mass and acceleration = 0.5 g: Max. power extracted = 0.678 mW (according the Jacobi's law Fig. 4(a))*

ii. *Six Thermal Electric generators as shown in block ② of Fig. 1 with efficient thermoelectric effect and 17 P&N stack for every single TEG. One example of maximum generation of power with a temperature gradient of 15°C is 0.562mW.*

   iii. *One thin-film amorphous silicon solar cell as shown in block ④ of Fig. 1 as the energy source for INDOOR artificial light energy harvesting with power density of 0.042 μW/$mm^2$ @ 200 lux(reference number AM-1801). The current/voltage ratio under this illumination level is 18.5uA @ 3.0 Vdc..*

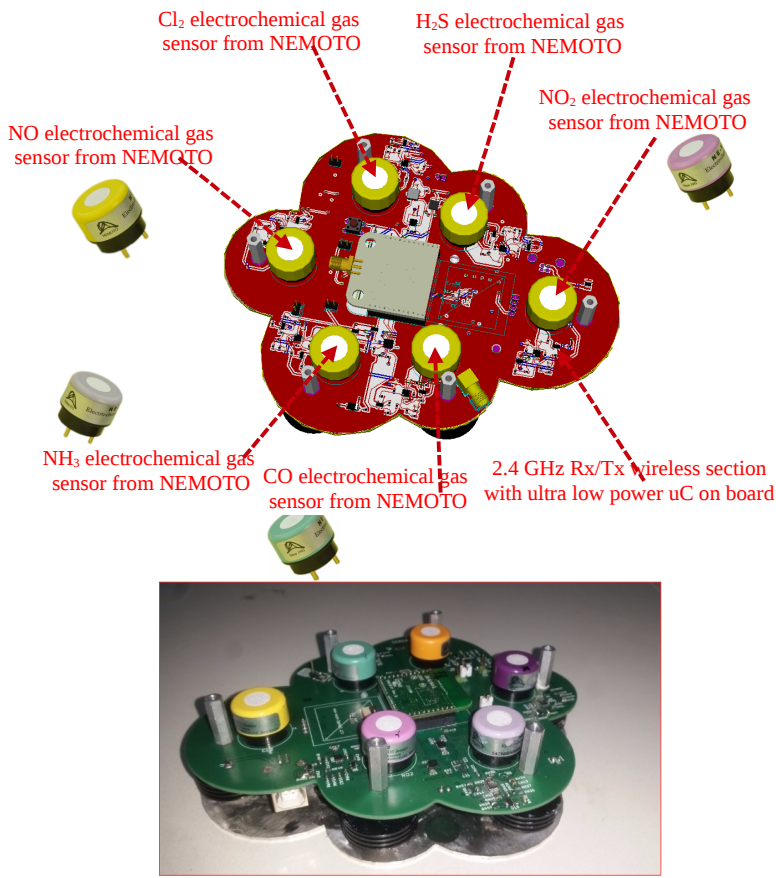

FIGURE 2 – 3D view of SERENO (PCB top view) and photos of the prototype where the 6 gas sensors are highlighted

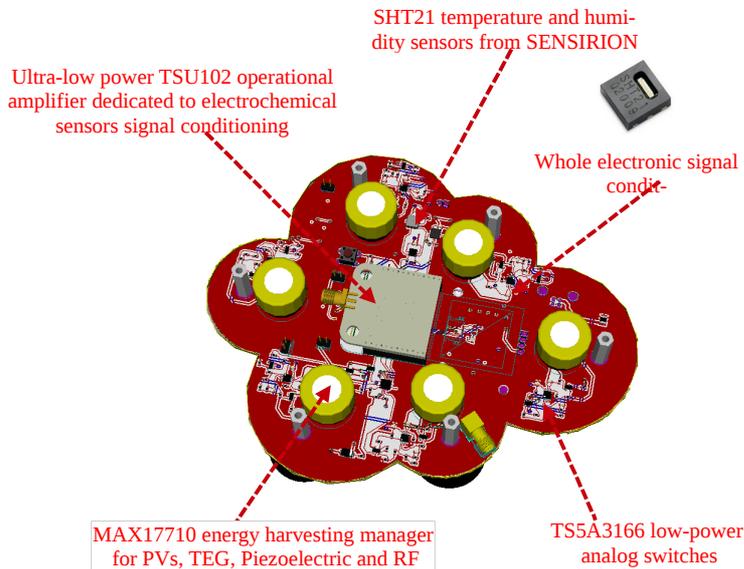

FIGURE 3 – 3D view of SERENO (PCB top view) where functional blocks are illustrated

In the next paragraphs the description of each section is given.

*Thin-film amorphous silicon solar cell (Fig. 1 block④) as energy source for OUTDOOR solar harvesting with power density of 1 µW/mm$^2$ @ 50 klux (AM-5904). The current/voltage ratio under this illumination level is 4.5mA @ 5.0 Vdc.*

  iv. One Radio Frequency power source at 915 MHz (Fig. 1 block ⑤) based on the Powercast P2110 harvester receiver and RF to DC converter.

An energy harvesting module, the LTC3109 is highly-efficient, auto-polarity and ultra-low voltage step-up converter has been used. This integrated circuit(IC) is ideal for harvesting excessive energy from extremely low input voltage sources, such as TEGs section. The LTC3109 is work in a way that designed to use two small external step-up transformers (1:100) to create an ultra-low input voltage DC/DC step-up converter.

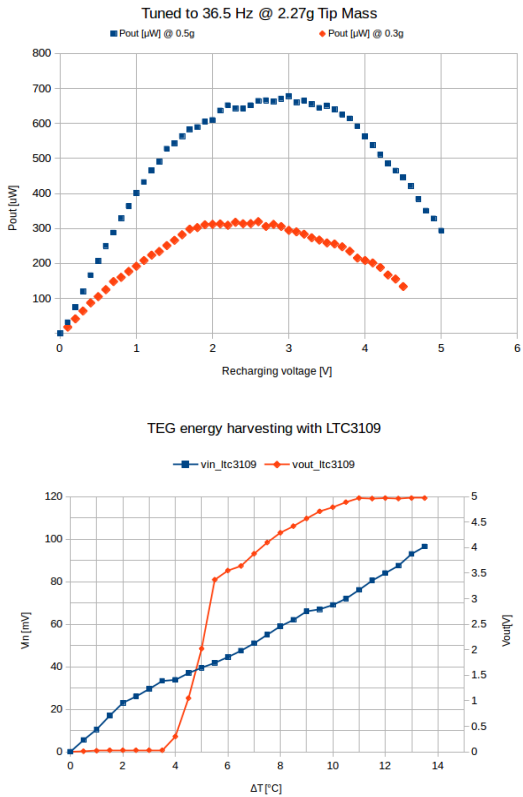

FIGURE 4(a) – Thermal gradient and vibration harvested power

### C. The Board equipped With Sensors

As already discussed, the fundamental purpose of the present study is battery-less autonomous operations. Therefore, 1mAh solid state battery, being charged from the energy section at 4.1 dc voltage, has been used as energy storage. The digital sensors of temperature and humidity, with the micro-controller and the RF module are powered at 3.3 dc voltage. The on board 10-bit A/D converter (SAR) convert accurate measurements of analog data of electrochemical sensors. A conditioning circuit for the gas sensors must be used. The signals we are getting directly from sensors are in the form of current, and it's inevitable to convert these signals into voltage for the ADC of the micro-controller. According to the data-sheet codes of the sensors the conditioning circuit has been designed using the ST TSU102 OP-AMPS. The major characteristics that makes this element suitable is supply current less than 600 nA at 3.3 $V_{dc}$ on each channel.

### D. Data transmission protocol

Wireless communication at 2.4 GHz is essential to send data to the main controller, which gathers the data from the sensors. This is the most power consuming operation of SERENO tasks. The design of this system is energy efficient since regular operations required that energy should be applied to transmission module only when it is necessary. Moreover it is necessary to have a robust and power-saving transmission protocol, which helps in achieving low energy consumption and will ensure data integrity in noisy environment. The RF module works with the IEEE 802.15.4 standard that uses the spread spectrum coding technique, in noisy and destructive channels it gives better performance.

### IV. APPROACHES AND OUTCOMES

Four physical hardware prototypes of SERENO have been mounted and tested in real environment conditions, to evaluate their performance and functionality in real application deployment. Each SERENO board has been supplied with a harvesting section, this harvesting part can recover energy from the photo-voltaic indoor/outdoor solar cell panel, a piezoelectric module, six thermoelectric modules and a radio frequency harvester at 915 MHz.

All the sources load at the same time a 1 mAh organic solid state battery that for each test begins with an energy level equal to zero. An array of sensors have been used which includes six gas sensors ($NH_3$, $Cl_2$, CO, $NO_2$, NO and $H_2S$).

### A. Techniques

At first, analysis were conducted on every single sensor seated on the board to characterize how the sensor acknowledges to various gases concentration at 25°C in order to find out the accuracy and cross-sensitivity. Fig. 5(a), display the devices under test with gas cylinders using accurate gas concentrations.

A poly-controls machine have been used for electronic control of gas mass flow and a climatic chamber for temperature analysis. The electrochemical sensors NE4-CO (current≈ 70 nA per ppm), NE4-$NO_2$ (current ≈ 600 nA per ppm), NE4-$H_2$S-100 (current≈700nA per ppm), NE4-$NH_3$ (current of ≈ 40 nA per ppm), NE4-NO (current ≈ 400 nA per ppm), NE4-$CL_2$ (current ≈ 600 nA per ppm) were tested on each of four boards mounted. Fig. 5(b), shows direct accuracy of about 3%FS (full scale) for CO, an accuracy of about 5%FS for $H_2S$, an accuracy of about 10%FS for NH3, an accuracy of about 5%FS for $NO_2$, an accuracy of about 5%FS for $Cl_2$ and an accuracy of about 5%FS for NO measures.

To compensate the slight drift in output of sensors due to temperature, by hardware NTC resistor and in software using a polynomial function according to NEMOTO sensors data-sheets of fourth order, has been applied, and the measured value is stable at various temperature readings (Fig. 5(c) and 5(d)). Regarding the gas sensors the studies have been conducted, in a tailored test chamber in the Biomedical lab at Qatar University (Qatar), as shown in Fig. 5 (a). To maintain the required gas concentration a mass flow controller has been used inside the chamber from reference cylinders with definitive known concentration (e.g. 200 ± 0.26% ppm for CO). The data sensed by sensors of SERENOs are sent to a receiver gateway via 2.4GHz wireless line, that performs data processing operations and send further through Ethernet connection.

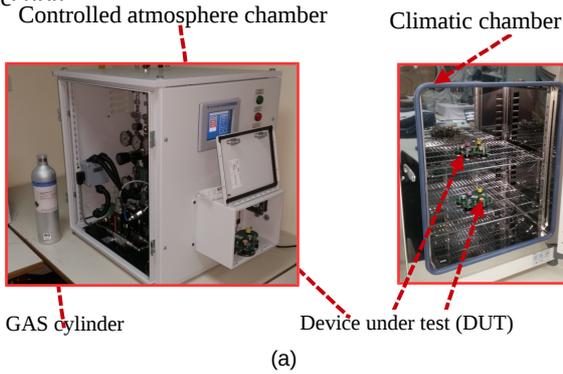

(a)

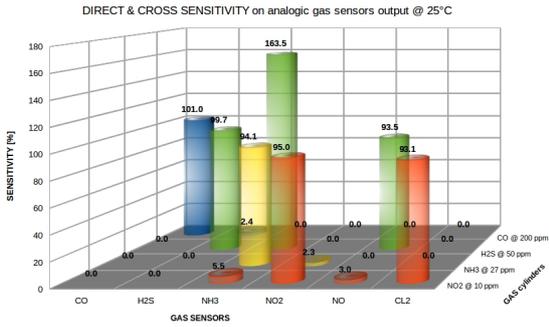

(b)

FIGURE 5 (a) and (b) – Devices under test with controlled atmosphere chamber and climatic chamber and final cross-sensitivity result

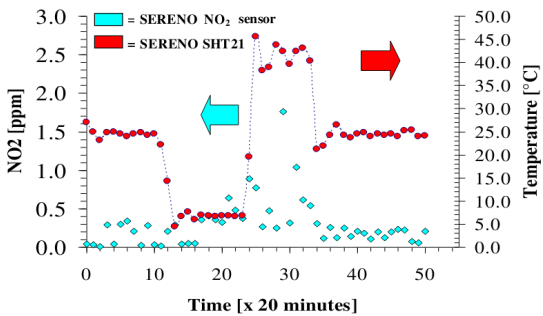

FIGURE 5 (c) – Final temperature drift after hardware (NTC) and software (4° order polynomial function) compensation techniques

### B. Measurement results

Fig. 5 (d) is an example of a measurement campaign that has been taken in Doha with a device placed inside a car dashboard, in the middle of the traffic jam, arranged in a day of normal urban side road, As we can see, the level of CO reaching the maximum value to 30 ppm. An influence of this reading for several number of hours in a day, can cause pestilent health effects by reducing oxygen delivery to the body's units and muscle tissues. However, the average value shown in the graph of Fig. 5 (d), is around 10 ppm. The path on the graph in red is marked by an autonomous GPS tracker.

The temperature graph points out peaks around 40°C during the first days of September. Humidity graph also shows value over 90% which are typical during that period of time in Doha.

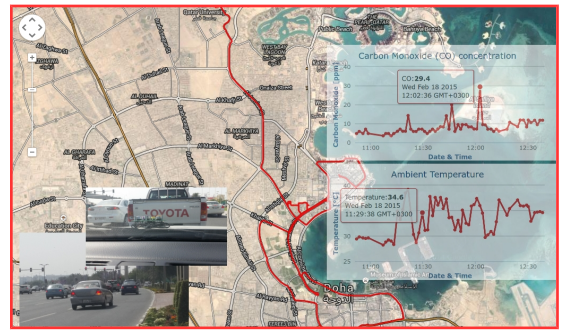

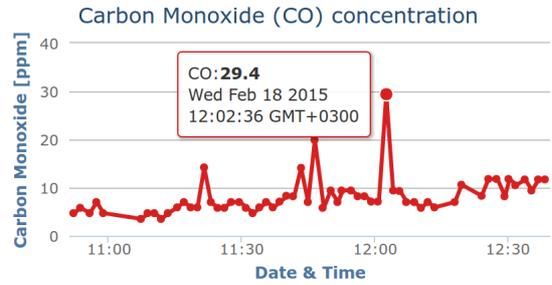

FIGURE 5 (d) – An example of measurement campaign in the traffic jam in Doha

| Operation | Energy per day (J/d) |
|---|---|
| Sensor warm-up (1 minute every 15 minutes) (115uA @ 3.7V) | 2.451 |
| Data Transmission and processing (4 seconds every 15 minutes) (25mA @ 3.7V) | 35.520 |
| Sleep Mode (14 minutes every 15 minutes) (3uA @ 3.7V) | 0.895 |
| **Total System Energy Budget Required by SERENO prototype per day** | **38.866** |
| **Total Energy Recovered per Day Under the Following Conditions:** | **39.529** |
| 1- Energy from Internal Artificial Light (office average: 8 hrs @ 200 lux) (3.0V @ 18.5µA) | 1.279 |
| 2- Energy from External light (sun average: 15min @ 50 klux) (5.0V @ 4.5mA) | 20.25 |
| 3- Energy from Temperature gradient through a window : 10 hrs with 6 °C (5 V @ 100 µA) and 14 Hrs with 0°C (non-operative) | 16.2 |
| 4- RF Energy @ 915 MHz: 3 W EIRP (transmitted) with SERENO positioned at 5 m (24 hrs operative) | 1.8 |
| *The power recovered [W] is been measured directly on the solid state cell so it include the conversion losses.* | |

Table 1 – Energy allocation per day under different conditions

### C. Energy harvesting evaluation

The final and third study about the system is the evaluation of power consumption Vs perpetual functionality evaluated. Experiments were conducted in indoor climate, where the designed system has been mounted on a glass of window (Fig. 6). The results of the energy obtained by the harvesting section and budget power consumption are shown in Table1. The sources from which energy being harvested were RF, internal/external light and thermoelectric TEGs. The harvesting source which was not utilized was piezoelectric generator in case of harvesting from vibrations, presume too difficult and uncertain to create a theoretical model.

As the test was in progress, a permanent back light source at 200 Lux has been focused in opposite of the device in order to simulate a real normal working day. To simulate the thermal gradient experiment, there should be at least 6°C thermal gradient difference between inner and outer surface, which is necessary value in order to collect energy from TEGs.

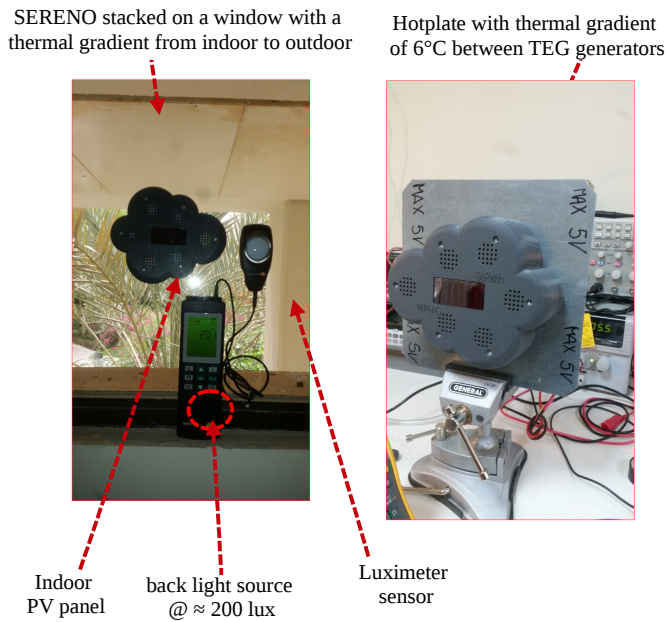

FIGURE 6–Devices under test for the energy harvesting section

V. CONCLUSION

In this work, we have presented a multi-parametric sensing platform called SERENO (SEnsor REceiver NOde) completely powered from the environmental sources, which shows a new approach to air quality monitoring. SERENO board design is able to manage intelligently energy transfer without human intervention, allowing its deployment in indoor and outdoor applications. The conducted experiments demonstrates that the described platform was able to operate as an air quality monitor in a "set and forget" scenario using a mesh network topology to cover large area for its deployment and also the system scalable and expandable without any further costs or any alteration in already deployed system.

VI. ACKNOWLEDGMENTS

This publication was made possible by NPRP grant # 6-203-2-086 from the Qatar National Research Fund. The statements made herein are solely the responsibility of the authors.

REFERENCES

[1] "7 million premature deaths annually linked to air pollution." http://www.who.int/mediacentre/news/releases/2014/air-pollution/en/.

[2] C. K. Chau, W. K. Hui, and M. S. Tse, "Evaluation of health benefits for improving indoor air quality in work place", *Environ. Int.*, vol. 33, no. 2, 2007, pp. 186–198.

[3] W. S. Cain, J. M. Samet, and M. J. Hodgson, "The quest for negligible health risks from indoor air", ASHRAE J., vol. 37, no. 7, p. 38-44, 1995.

[4] G. Hoek, B. Brunekreef, S. Goldbohm, P. Fischer, and P. A. van den Brandt, "Association between mortality and indicators of traffic-related air pollution in the Netherlands: A cohort study", Lancet vol. 360, no. 9341, pp. 1184–1185, Oct. 2002.

[5] Kjellstrom, T.E.; Neller, A. Simpson and R. W.: "Air Pollution and Its Health Impacts: The Changing Panorama", Medical Journal of Australia 2002, 177 (11-12), pp. 604-608.

[6] Richards M., Ghanem M., Osmond M., Guo Y. and Hassard J.: "Grid-based Analysis of Air Pollution Data", Ecological Modelling 2006, 194, pp. 274-286.

[7] S. Abraham and L. Xinrong, "A Cost-effective Wireless Sensor Network System for Indoor Air Quality Monitoring Applications," in Procedia Computer Science, vol. 34, 2014, pp. 165-171

[8] "Impact Analysis," European Commission, 2013.

[9] Chaichana Chaiwatpongsakorn, Mingming Lu, Tim C. Keener and Soon-Jai Khang: "The Deployment of Carbon Monoxide Wireless Sensor Network (CO-WSN) for Ambient Air Monitoring", Int. J. Environ. Res. Public Health 2014, 11, 6246-6264.

[10] S. Devarakonda, P. Sevusu, H. Liu, R. Liu, L. Iftode, B. Nath: "Real-time air quality monitoring through mobile sensing in metropolitan areas", In Proc. of UrbComp '13. ACM, New York.

[11] Yun Cheng, Xiucheng Li, Zhijun Li, Shouxu Jiang, Yilong Li, Ji Jia and Xiaofan Jiang: "AirCloud: A Cloud-based Air-Quality Monitoring System for Everyone", SenSys '14 Proc. of the 12th ACM Conference of Embedded Network Sensor Systems, pp. 251-265.

[12] Yifei Jiang, Kun Li, Lei Tian, Ricardo Piedrahita, Xiang Yun, Omkar Mansata, Qin Lv, Robert P. Dick, Michael Hannigan and Li Shang: "MAQS: a personalized mobile sensing system for indoor air quality monitoring", Proc. of the 13th international conference on Ubiquitous computing, September 17-21, 2011, Beijing, China.

[13] M. Yajie, M. Richards, M. Ghanem, Y. Guo and J. Hassard, "Air Pollution Monitoring and Mining Based on Sensor Grid in London", Sensors, Vol. 8, No.6, 2008, pp. 3601-3623.

[14] J. Kim, C. Chu, S. Shin: "ISSAQ: An Integrated Sensing Systems for Real-Time Indoor Air Quality Monitoring", IEEE Sensors Journal 14 (12),pp. 4230 – 4244.

[15] Honicky, R., Brewer, E.A., Paulos, E., White, R.: "N-smarts: networked suite of mobile atmospheric real-time sensors", In Proceedings of the 2nd ACM SIGCOMM Workshop on Networked Systems for Developing Regions, Seattle, WA, USA; ACM: Seattle, WA, USA, 2008, pp. 25-30.

[16] Octavian Postolache, José Miguel Pereira, Pedro Silva Girão, António Almeida Monteiro, "Greenhouse Environment: Air and Water Monitoring Smart Sensing Technology for Agriculture and Environmental Monitoring", Lecture Notes in Electrical Engineering Volume 146, 2012, pp 81-102

[17] "Electrochemical sensors for environment monitoring: A review of recent technology", (Mar. 2005) Department of Chemistry and Biochemistry New Mexico State University. Available on: http://clu-in.org/download/char/sensr_ec.pdf

[18] J.Wang, "Modified electrodes for electrochemical sensors", Electroanalysis, vol. 3, nos. 4–5, 1991, pp. 255-259.

[19] U. Yogeswaran and S. M. Chen, "A review on the electrochemical sensors and biosensors composed of nanowires as sensing material," Sensors, vol. 8, no. 1, pp. 290–313, 2008.

[20] Alphasense sensors data sheet, Shielding toxic sensors from electromagnetic interface, no. 12. Great Notley, U.K., Alphasense Ltd., Sensor Technology House, 2005.

[21] E. Bakker and Y. Qin, "Electrochemical sensors", Anal. Chem., vol. 78, no. 12, 2006, pp. 3965–3984.

[22] F. Estini: "Electrochemical sensor: kit comprising said sensor and processor for the production", 2008 [Online]. Available: http://www.faqs.org/patents/app/20080251393

[23] Alphasense sensors data sheet, how electrochemical sensors work, no. 12. Great Notley, U.K., Alphasense Ltd.Sensor Technology House, 2005.

[24] D. E. Garcia, T. H. Chen, F. Wei, and C. M. Ho, "A parametric design study of an electrochemical sensors", J. Assoc. Lab. Autom., vol. 15, no. 3, pp. 179–188, Jun. 2010.

[25] U.S. Environmental Protection Agency. Air Pollution Monitoring [Online]. Available: http://www.epa.gov/oar/oaqps/montring.html

[26] Nemoto Chemical Sensors Catalogue, Nemoto & Co., Tokyo, Japan. [Online]. Available: http://www.nemoto.eu/products.html